\documentclass[twoside,slac_one]{revtex4}
\usepackage{graphicx}
\usepackage{fancyhdr}
\usepackage{amsmath} 
\usepackage{bm}
\usepackage{amsxtra}
\usepackage{amssymb}
\usepackage{amsthm}
\usepackage{latexsym}
\usepackage{lscape}
\usepackage{pennames}

\newcommand{\tev}{\ensuremath{\mathrm{\,Te\kern -0.1em V}}}
\newcommand{\gev}{\ensuremath{\mathrm{\,Ge\kern -0.1em V}}}
\newcommand{\mev}{\ensuremath{\mathrm{\,Me\kern -0.1em V}}}
\newcommand{\kev}{\ensuremath{\mathrm{\,ke\kern -0.1em V}}}
\newcommand{\ev}{\ensuremath{\mathrm{\,e\kern -0.1em V}}}
\newcommand{\gevc}{\ensuremath{{\mathrm{\,Ge\kern -0.1em V\!/}c}}}
\newcommand{\mevc}{\ensuremath{{\mathrm{\,Me\kern -0.1em V\!/}c}}}
\newcommand{\gevcc}{\ensuremath{{\mathrm{\,Ge\kern -0.1em V\!/}c^2}}}
\newcommand{\mevcc}{\ensuremath{{\mathrm{\,Me\kern -0.1em V\!/}c^2}}}

\begin{document}

\title{Soft QCD Results from the CMS Experiment}

%

\author{Dayong Wang, for the CMS Collaboration}
\affiliation{Department of Physics, University of Florida,
  Gainesville, FL 32611, USA}

\begin{abstract}

  Recent CMS soft QCD results in proton-proton collisions at three LHC
  center-of-mass energies are highlighted. The properties of minimum
  bias events such as charged particle transverse momentum spectra,
  event-by-event multiplicity distributions, the production of the
  strange particles are measured. Particle correlations, such as long-
  and short-range angular correlation as well as Bose-Einstein
  correlation are studied. Characteristics of the underlying event
  and comparisons to MC tunes have been made.

\end{abstract}

\maketitle

\thispagestyle{fancy}


\section{Introduction}

In a high energy proton-proton collision environment such as the
Large Hadron Collider (LHC), the majority of the collisions are quite
soft, without hard parton scatterings. In an event that does have
parton scatterings with high-$Q^2$ transfer, the outgoing partons have
to go through the parton showering and hadronization process. In addition,
there will also be the accompanying Underlying Events (UE), which are
mainly from Multiple Parton Interactions (MPI) and Beam-Beam Remnants.  

All those QCD processes are soft, which means no perturbative
predictions are available. The experiments usually have to rely on the
Monte Carlo (MC) descriptions to model them phenomenologically. The
models prior to LHC data were tuned on previous measurements. When
extrapolated to LHC energy, different models diverge at the
predictions.

Early LHC data collected by the Compact Muon Solenoid (CMS)
experiment \cite{cmsdetector} thus provide a unique chance to deepen
our knowledge on the soft QCD through the well designed
measurements. They are crucial for precision measurements of the
standard model processes and for new physics searches.

In addition, these measurements on the pp collisions will provide
necessary references for physics studies with heavy ion collisions at
CMS.

\section{Properties of the Minimum Bias Events}

In CMS, the minimum bias (MB) events are mainly triggered by the
designed triggers through coincidence of signals from monitoring
components of Beam Pickup Timing for experiments (BPTX) and Beam
Scintillator Counters (BSC). Further selections from information
recorded in other detectors are applied offline to select the events.
In order to extend the transverse momentum reach, some single jet
triggers based on signals from calorimeters are also used in the
analysis.

\begin{figure*}[!hbt]
\label{fig:mbptspectra}
\centering
\includegraphics[width=0.32\textwidth]{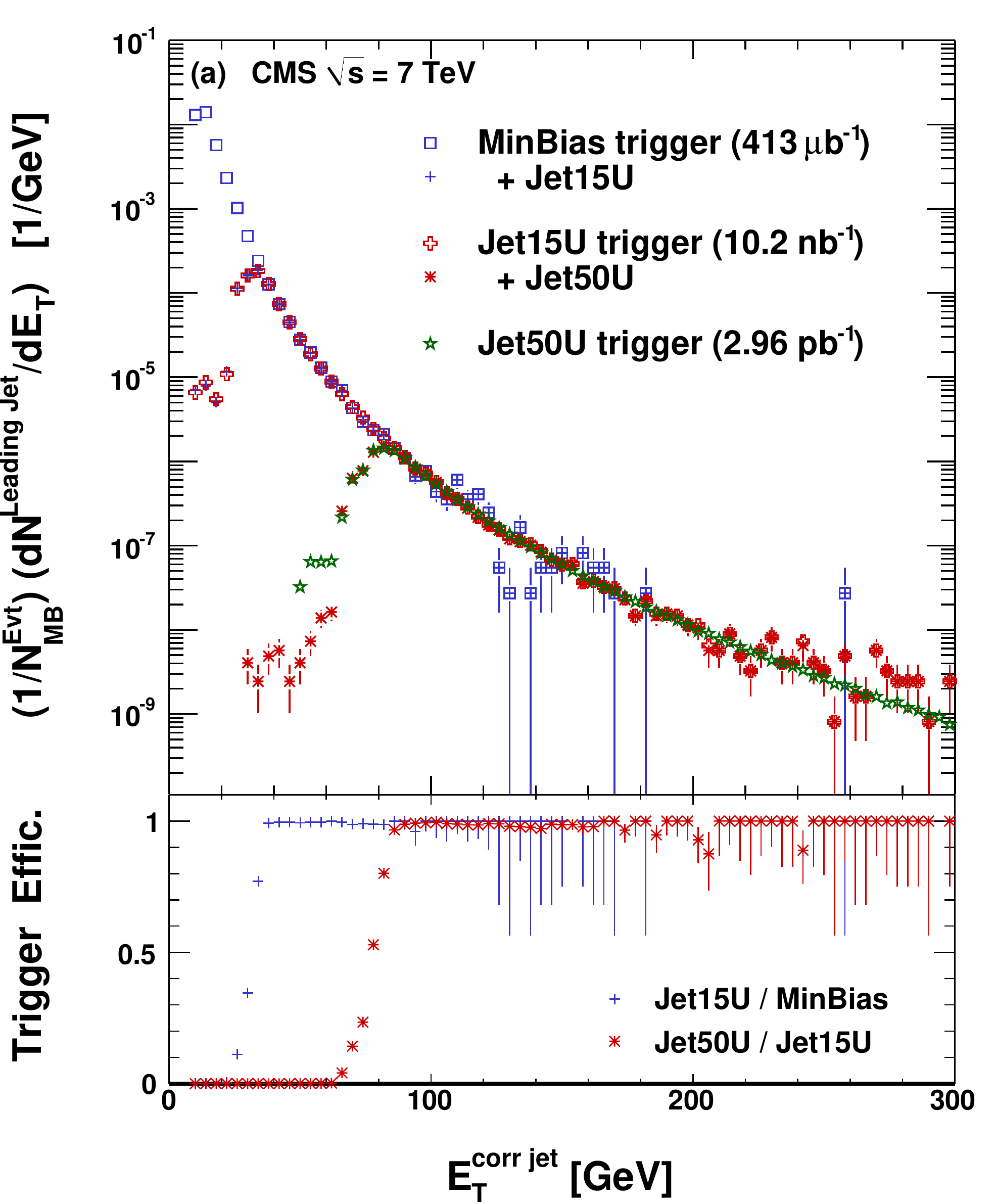}
\includegraphics[width=0.32\textwidth]{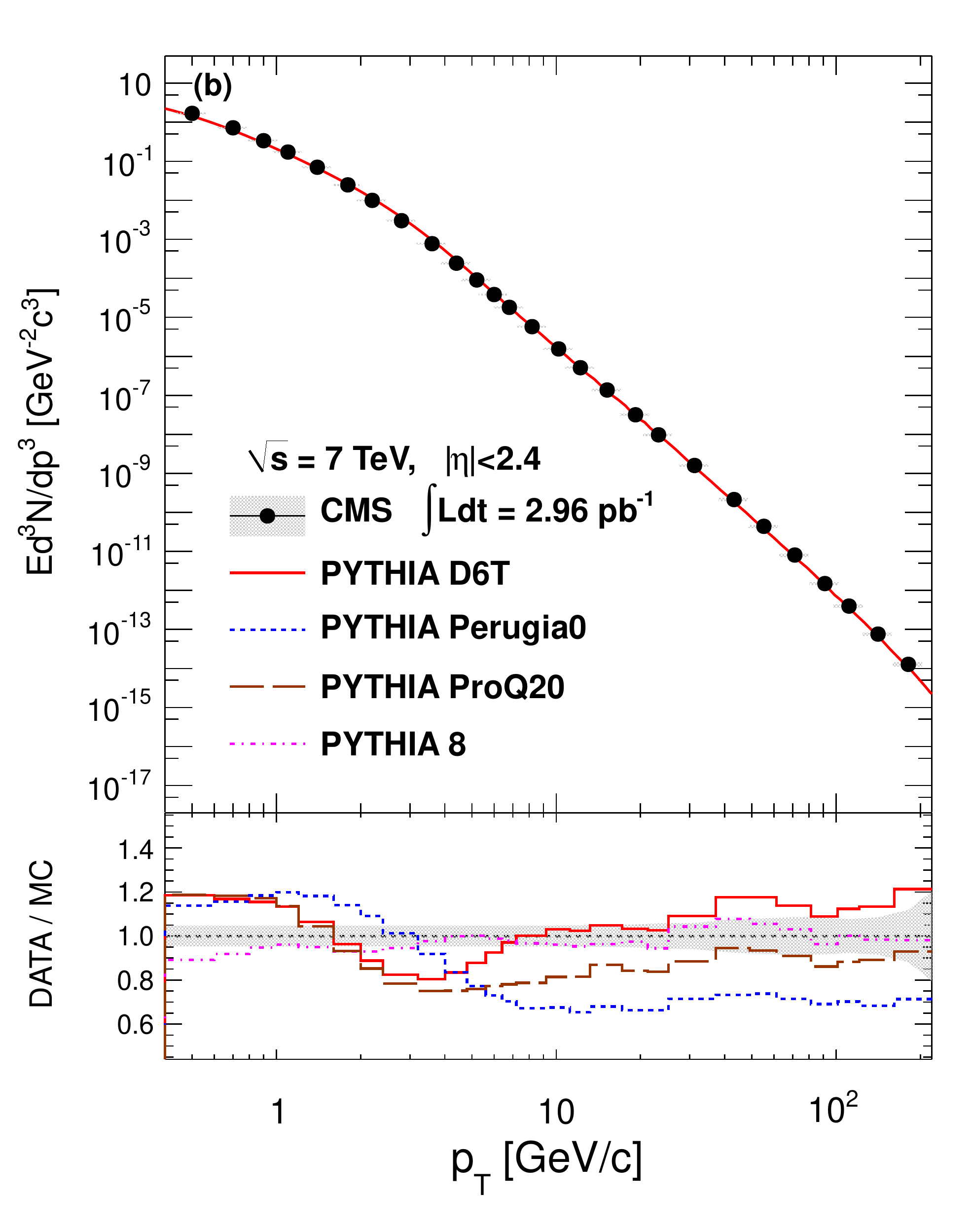}
\includegraphics[width=0.32\textwidth]{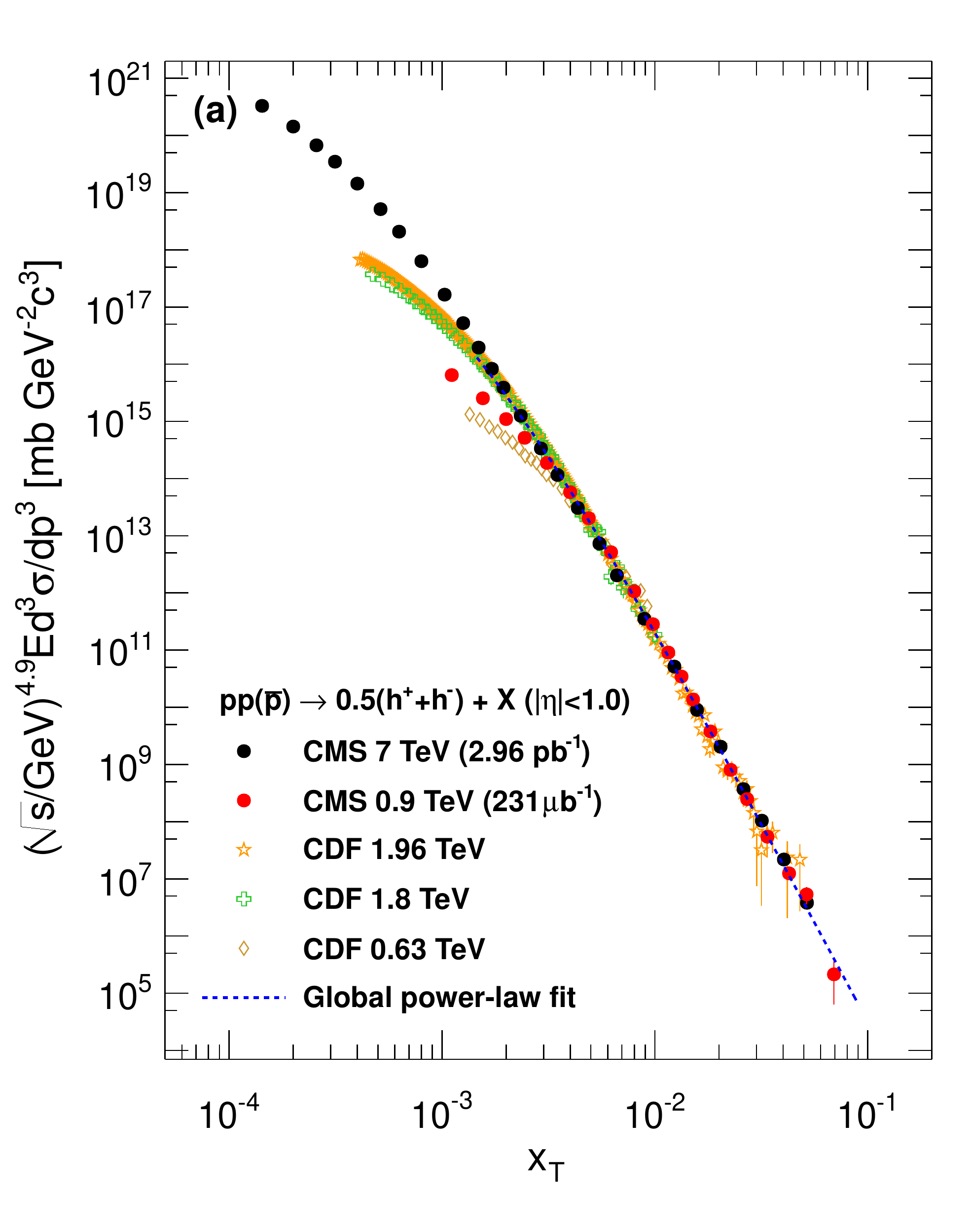}
\caption{Transverse momentum spectra of the charged particles in MB events.} \label{figure1}
\end{figure*}

The transverse momentum spectra of charged particles in MB events as
measured in CMS \cite{ptspectra} are shown in the
Fig. \ref{fig:mbptspectra}. The left plot shows the yield with respect
to the jet transverse momenta from different data sets with different
triggers, which are fully efficient above the trigger thresholds as
shown in the bottom panel. The middle plot compares the phase-space
invariant event yield from the data to various MC models. None of them
can describe the spectra perfectly, but Pythia8 is more compatible
than Pythia6 models. The right plot confirms the scaling behavior with
the variable $x_T=2p_T/\sqrt{s}$ from measurements performed by
different experiments at different energies.

\begin{figure}[!hbt]
  \centering
  \includegraphics[width=0.5\textwidth]{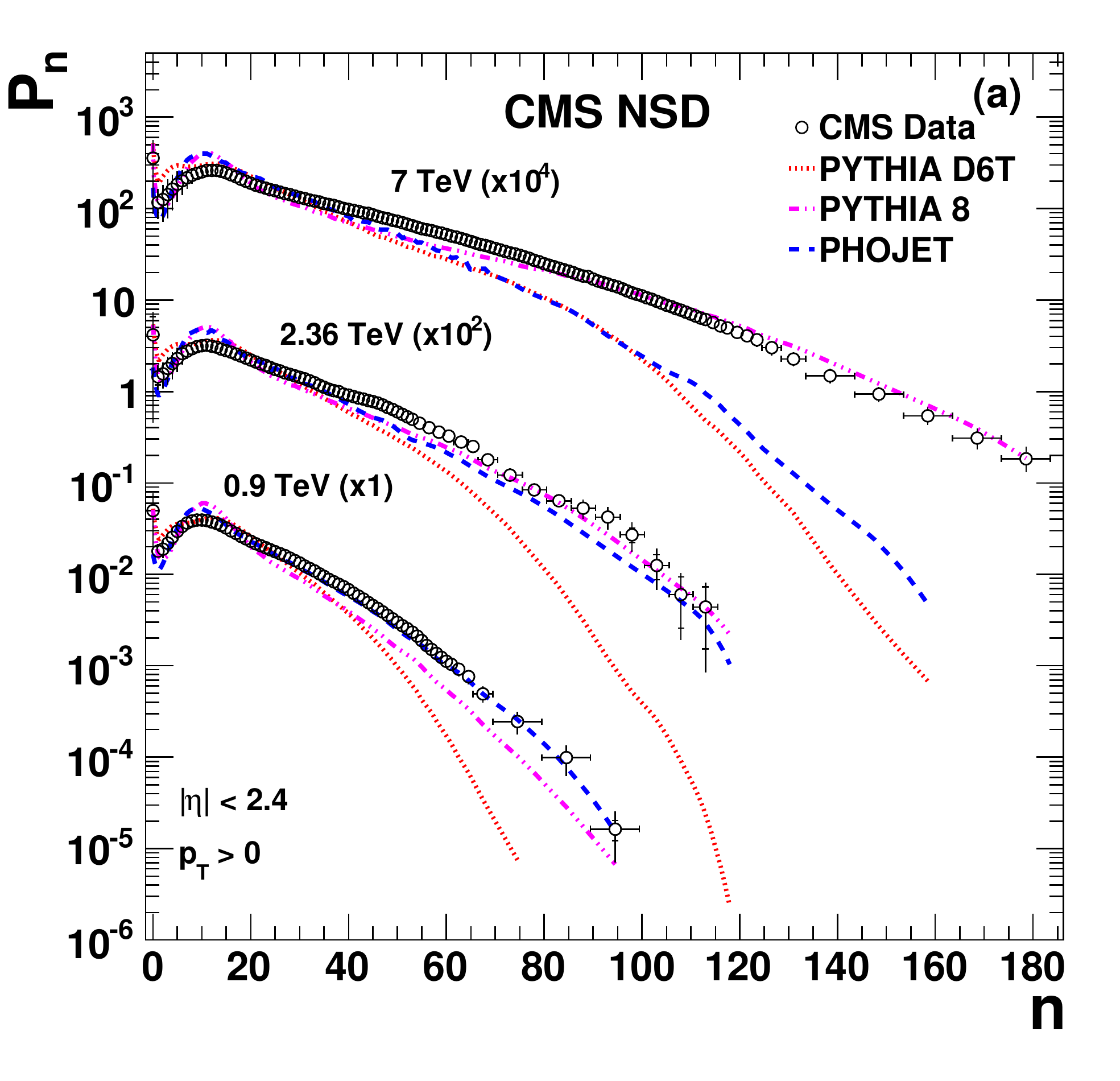}
  \caption{Charged particle multiplicities measured by CMS at three LHC energies.}
  \label{fig:mbmultiplicity}
\end{figure}

Charged particle multiplicities in MB events as measured by CMS
\cite{multiplicities} at three LHC energies (0.9 TeV, 2.36 TeV  and
7 TeV) are shown in Fig.  \ref{fig:mbmultiplicity}, with different
shifts in the scales for visibility. A large tail in multiplicities is
observed in 7 TeV, which can reach nearly 200 tracks. Most MC models
can not describe the multiplicities well in all energies for these
unprecedented measurements, especially in the large multiplicity
end. Further tunings are thus required, but Pythia8 shows better
agreements with data.

\begin{figure}[!hbt]
  \centering
\includegraphics[width=0.32\textwidth]{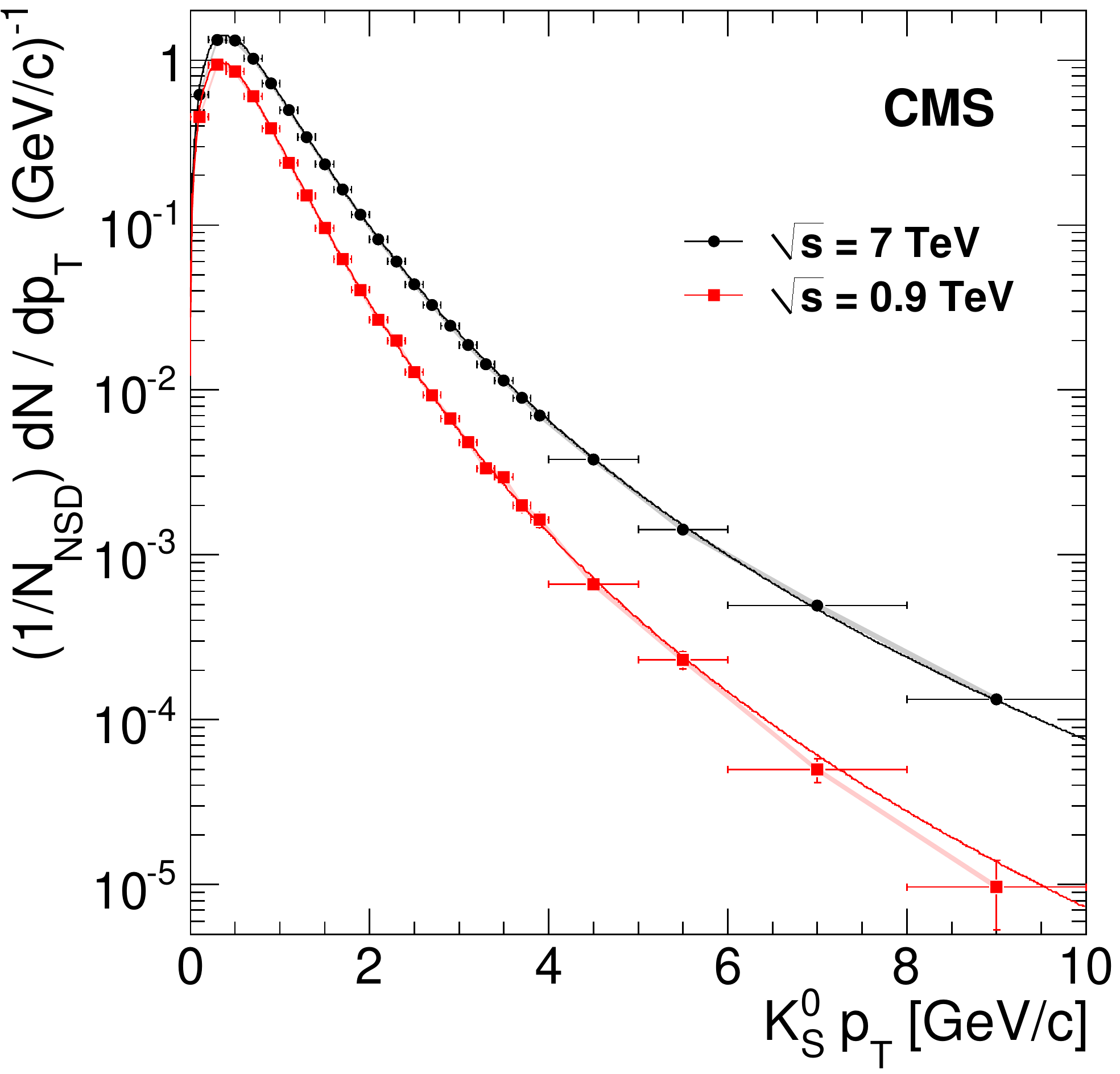}
\includegraphics[width=0.32\textwidth]{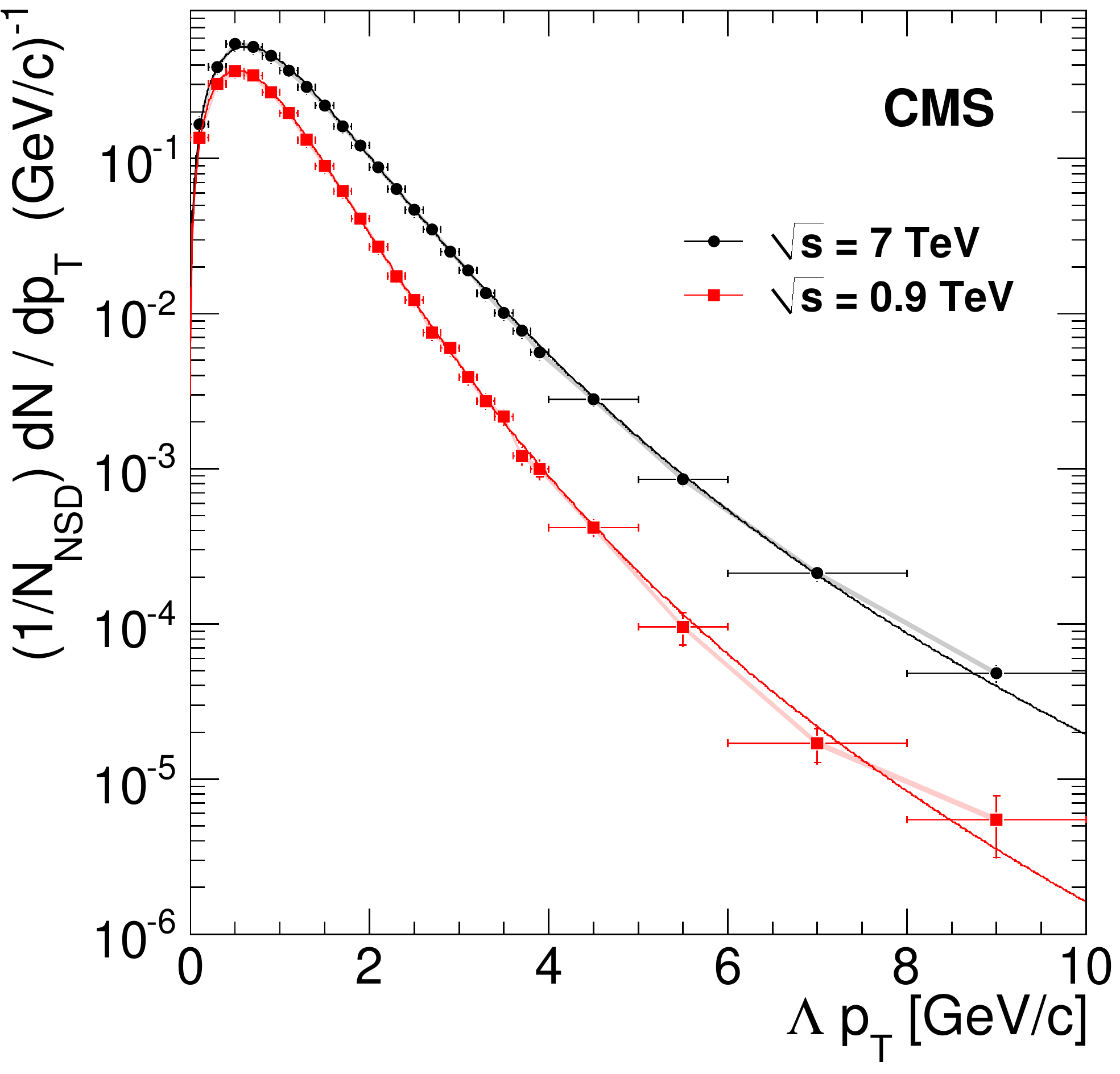}
\includegraphics[width=0.32\textwidth]{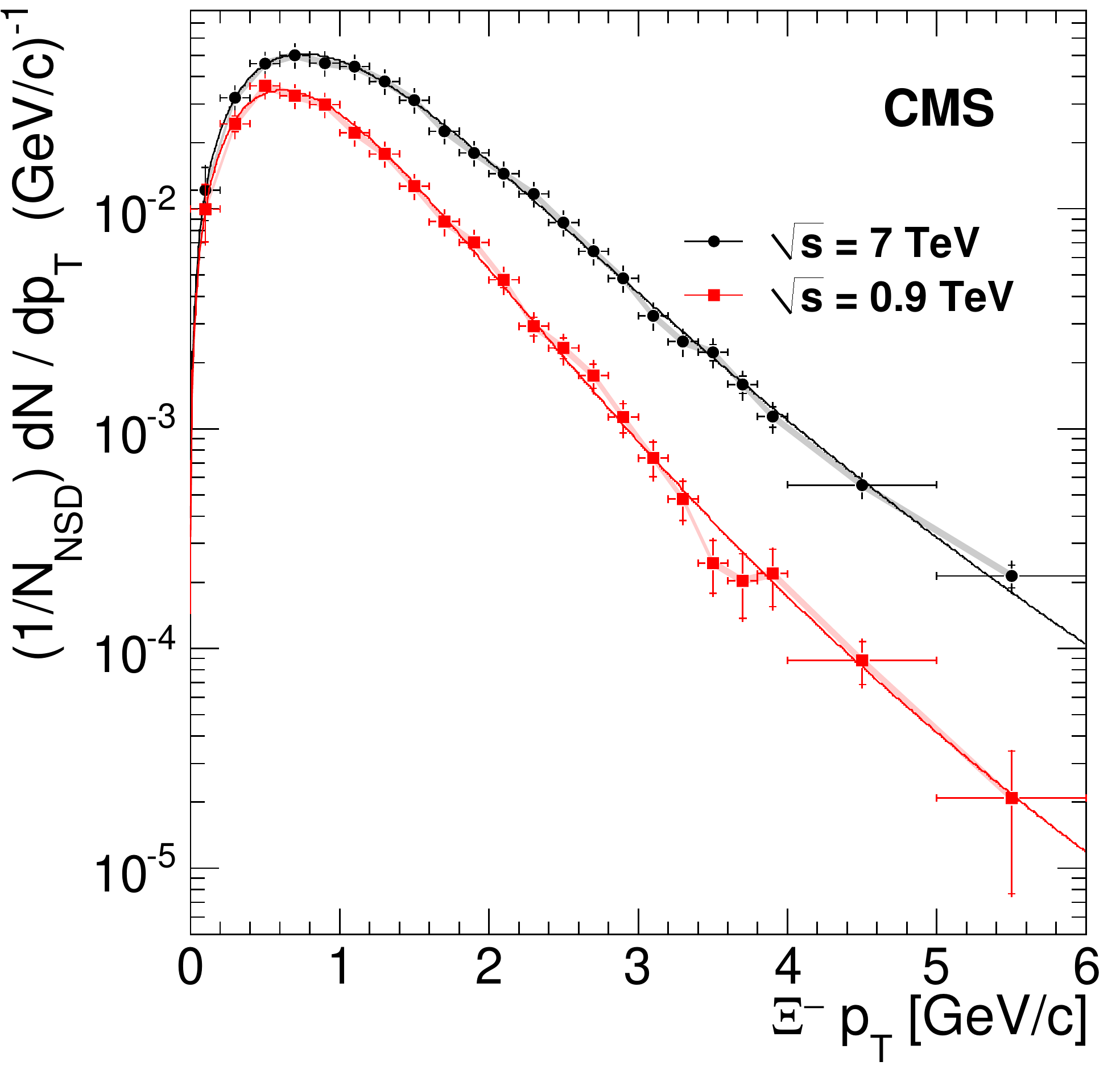}
  \caption{Distributions of strange particle yields at 0.9 and 7 TeV
    w.r.t. transverse momenta, normalized to total Non Single
    Diffractive (NSD) events.}
  \label{fig:strangeyieldvspt}
\end{figure}

Strange particle productions are also studies from MB events, at both
at 0.9 and 7 TeV, and compared \cite{strangeproduction}. In the MB
data sets, strange meson \PKzS \ and hyperons \PgL \ , \PgXm \ can be
reconstructed from their decay products well with very good mass
resolutions and low background contaminations. The masses and widths
are all consistent with PDG values. The differential event yields are
studied with respect to both strange particle transverse momentum 
and pseudorapidity (denoted as $\eta$).  For the $p_T$ spectra
shown in Fig.  \ref{fig:strangeyieldvspt}, the empirical fits can
describe the data well at both energies. With increase of the LHC
energy, the strange particle $p_T$ spectra all become harder. At 7
TeV, all the $p_T$ spectra are flatter at the middle where
exponential falling dominates. For \PKzS \ spectrum (left), it is also
flatter at the higher end where the negative power law dominates. This
is not as obvious as in the case of \PgL \ (middle) and \PgXm \ (right).

\begin{figure}[!hbt]
  \centering
\includegraphics[width=0.32\textwidth]{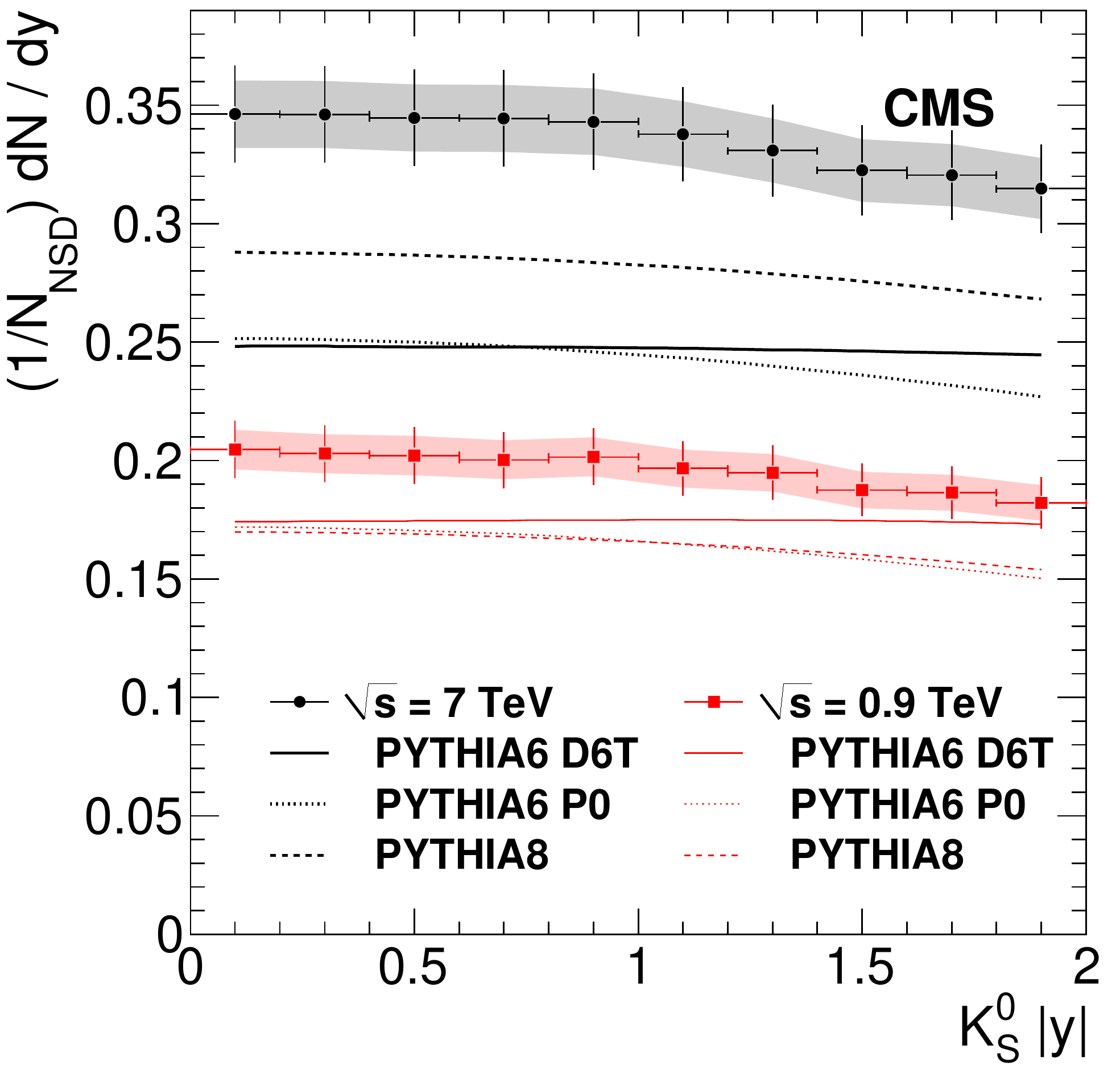}
\includegraphics[width=0.32\textwidth]{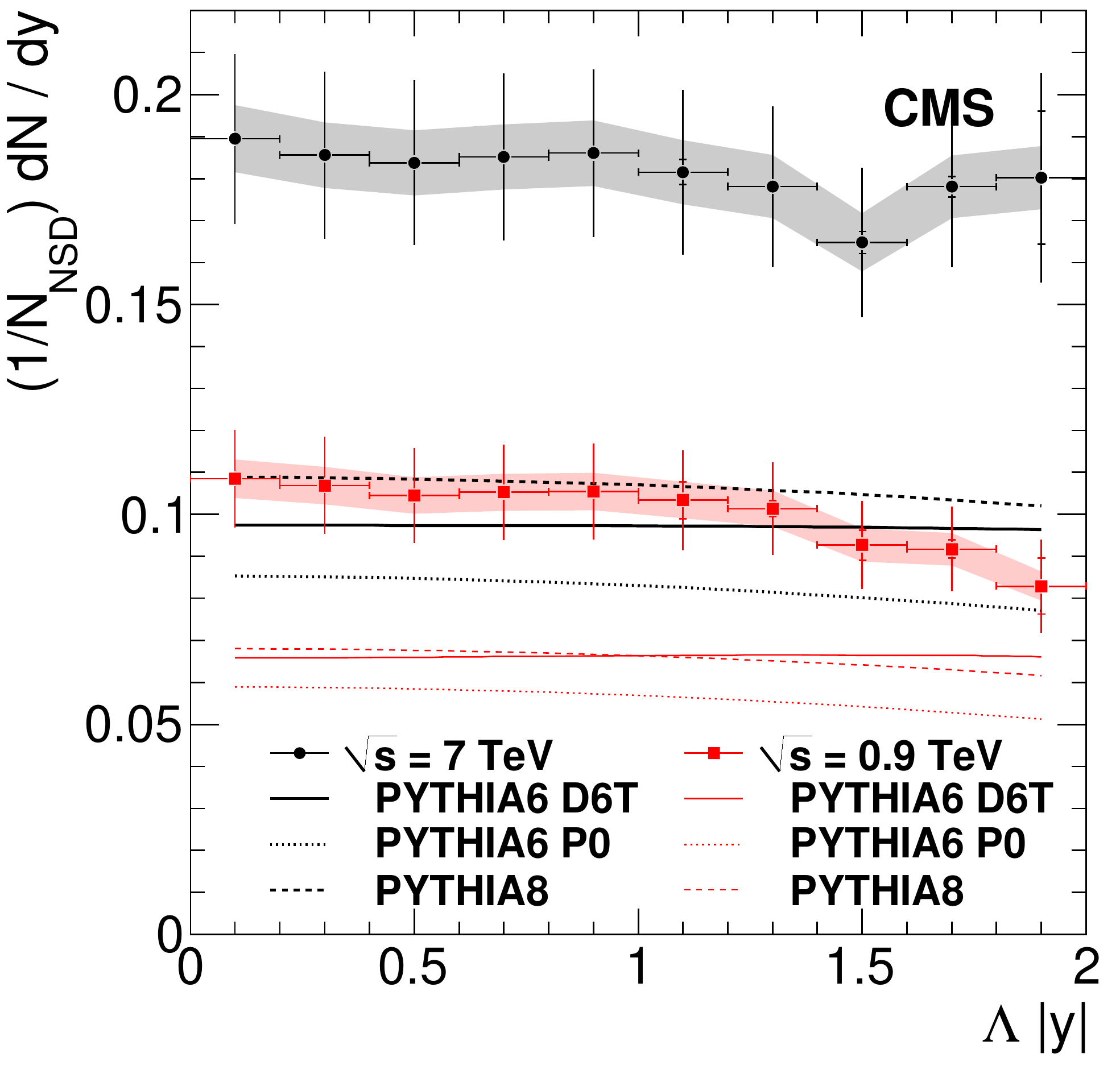}
\includegraphics[width=0.32\textwidth]{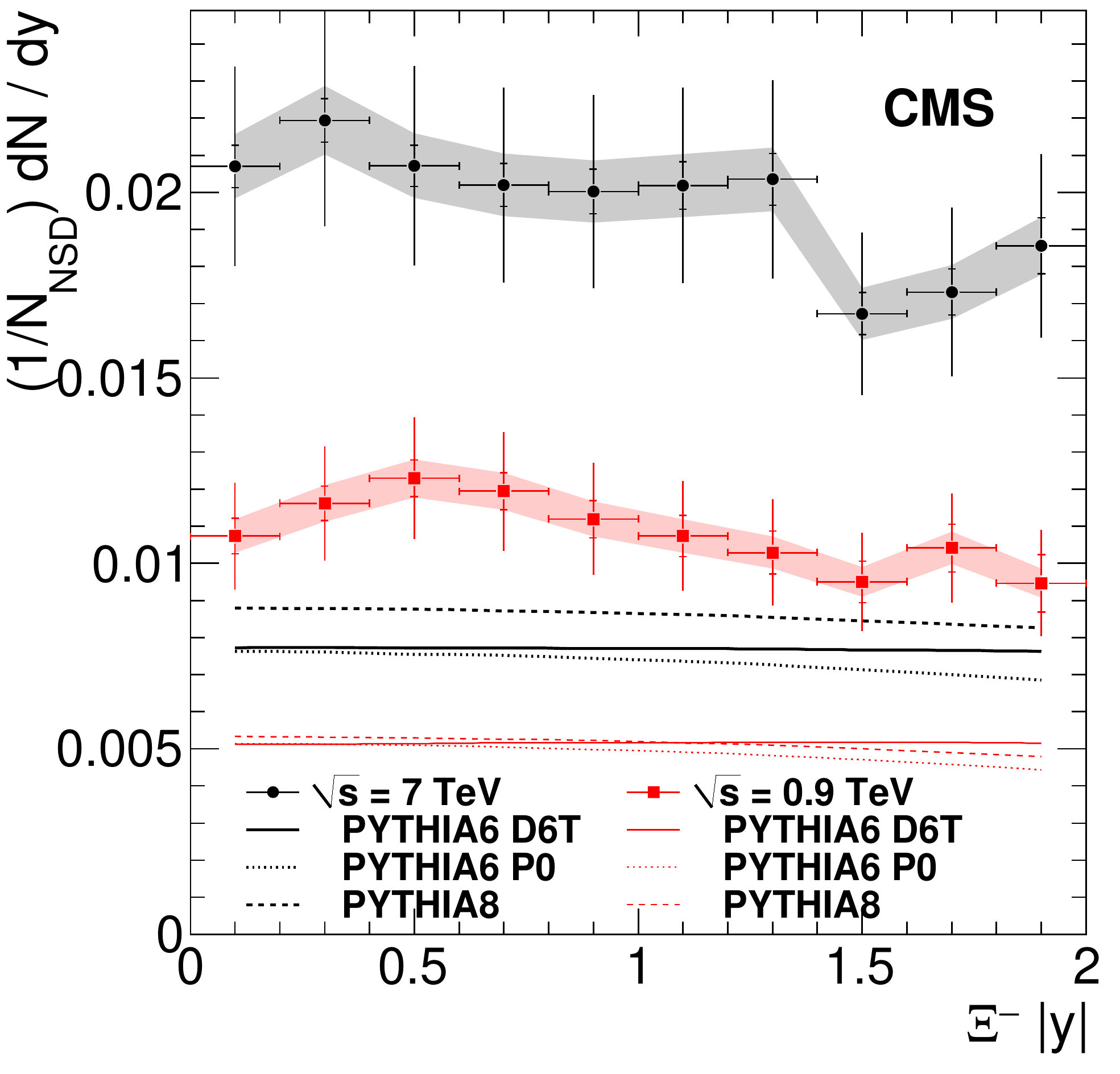}
\caption{Distributions of strange particle yields at 0.9 and 7 TeV
  w.r.t. pseudorapidities, normalized to total Non Single
  Diffractive (NSD) events. The inner vertical error bars  (when
  visible) show the statistical uncertainties, the outer the
  statistical and point-to-point systematic uncertainties summed in
  quadrature. The normalization uncertainty is shown as a band.}
  \label{fig:strangeyieldvseta}
\end{figure}

The $\eta$ distributios of strange particles are compared to
different MC models and tunes for both LHC energies, as shown in
Fig. \ref{fig:strangeyieldvseta}. From 0.9 TeV to 7 TeV, the strange
particle yields are doubled, with respect to total Non Single
Diffractive (NSD) events. All MC models show big amount of
underestimation and need further tuning to be used for future studies of
possible strange particle suppression.

\section{Measurements of Particle Correlations}

\subsection{Two-particle correlation in $\Delta \eta$ and $\Delta \phi$}
\label{sec:twoparticlecorr}

\begin{figure}[!hbt]
  \centering
\includegraphics[width=0.45\textwidth]{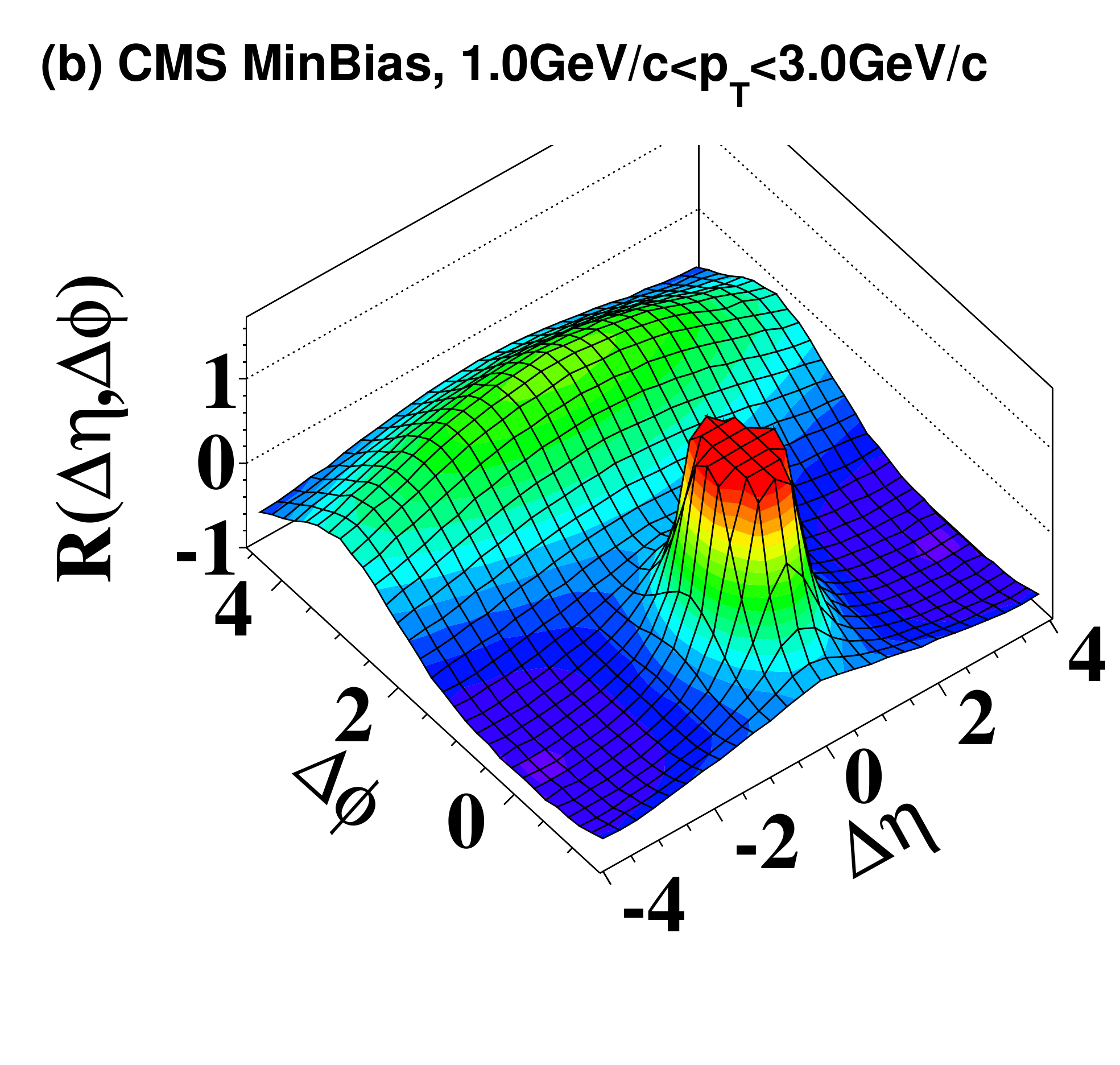}  
\includegraphics[width=0.45\textwidth]{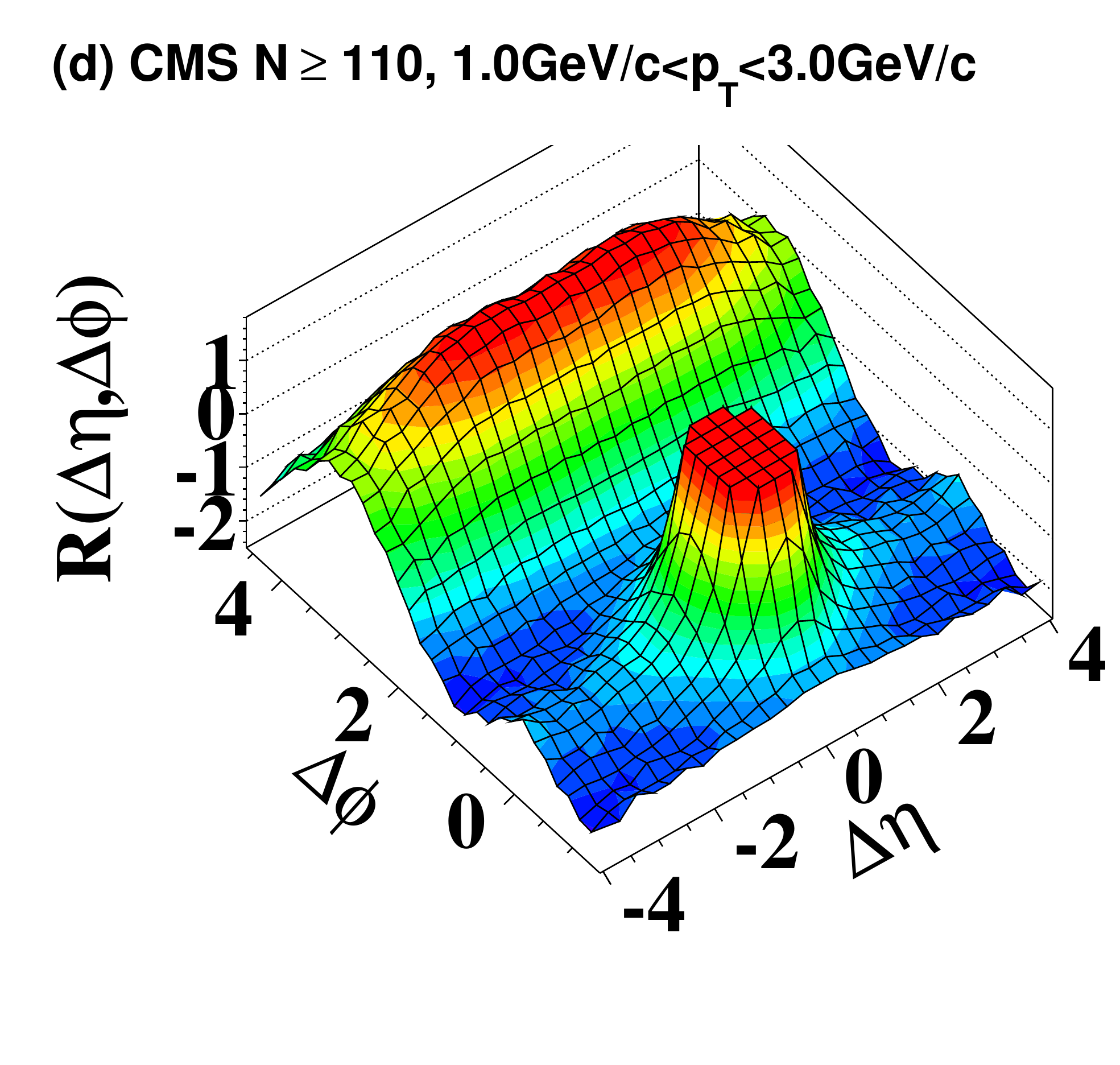}  
  \caption{Two-particle correlations observed at CMS in MinBias events
    and in high multiplicity events. Both plots shown are for
    particles in the intermediate transverse momenta range of $1.0 \gevc < p_T < 3.0 \gevc$.}
  \label{fig:twoparticlecorrelation}
\end{figure}

Two-particle correlation in $\Delta \eta$ and $\Delta \phi$ reflects
the underlying production mechanism. While the CMS measurements of
this correlation in the inclusive MinBias events at 0.9 and 7
TeV \cite{2partridge} have shown all known features as predicted by
MC models, the exclusive measurements with the events selected from
the specially designed high multiplicity triggers have shown some
unexpected features as shown in Fig.
\ref{fig:twoparticlecorrelation}.

The near side, long range, angular correlation is established at
high multiplicities. It is especially pronounced for the intermediate
transverse momenta range $1.0 \gevc < p_T < 3.0 \gevc$. This
feature could not be reproduced from MC simulations. Together with some
heavy ion observations, it may indicate the formation of new state of
matter in $pp$ collisions at this energy, so it provides a new test
ground for high density QCD physics.

\subsection{Bose-Einstein Correlation (BEC) measurements}
\label{sec:bec}

The production probability of the identical boson pairs with similar
momenta could be enhanced due to the Bose-Einstein
Correlation (BEC). The measurements of BEC can provide important
information for the size, shape, space-time development of the emitting
source.

CMS makes BEC measurements at 0.9 and 7 TeV and studies the
dependences of this enhancement on kinematic and topological features
of the events \cite{bec2011}. The ratio of the enhancement is
quantified with respect to reference samples with non interfering
boson pairs. Its dependence on the Lorentz-invariant variable
$Q=\sqrt{-(p1-p2)^2}$ is usually parameterized with several physical
parameters such as the correlation length, long distance correlations,
effective size of the emission region, etc.

\begin{figure}[!hbt]
  \centering
  \includegraphics[width=0.5\textwidth]{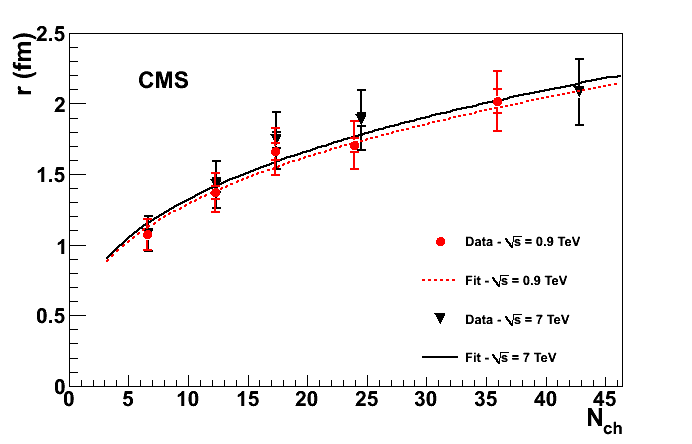}    
  \caption{Effective BEC emission size $r$  as a function of the charged-particle multiplicity in the event.}
  \label{fig:becsizevsn}
\end{figure}

Fig. \ref{fig:becsizevsn} clearly establishes the effective emission
size $r$ as a function of the charged-particle multiplicity in the
event. It confirms for the first time the growth of BEC effective
emission region with multiplicity which was ambiguously inferred from
previous measurements. This dependence can also account for the growth
with $\sqrt{s}$ very well.


\section{Underlying event (UE) measurements and MC Tunes}

The underlying event (UE) includes hadronic activity that is additional to those
from hard scattering processes. It arises mainly from multiple parton
scattering (MPI) and beam-beam remnants etc. It is most sensitive in
the transverse region, $60^\circ < |\Delta \phi|< 120^\circ$, as a
function of the $p_T$ of the leading track-jet.  CMS has performed the
fully corrected measurements of charged particles from UE of 0.9 TeV
and 7 TeV \cite{ue2010} \cite{ue2011}.

\begin{figure}[!hbt]
  \centering
  \includegraphics[width=0.32\textwidth]{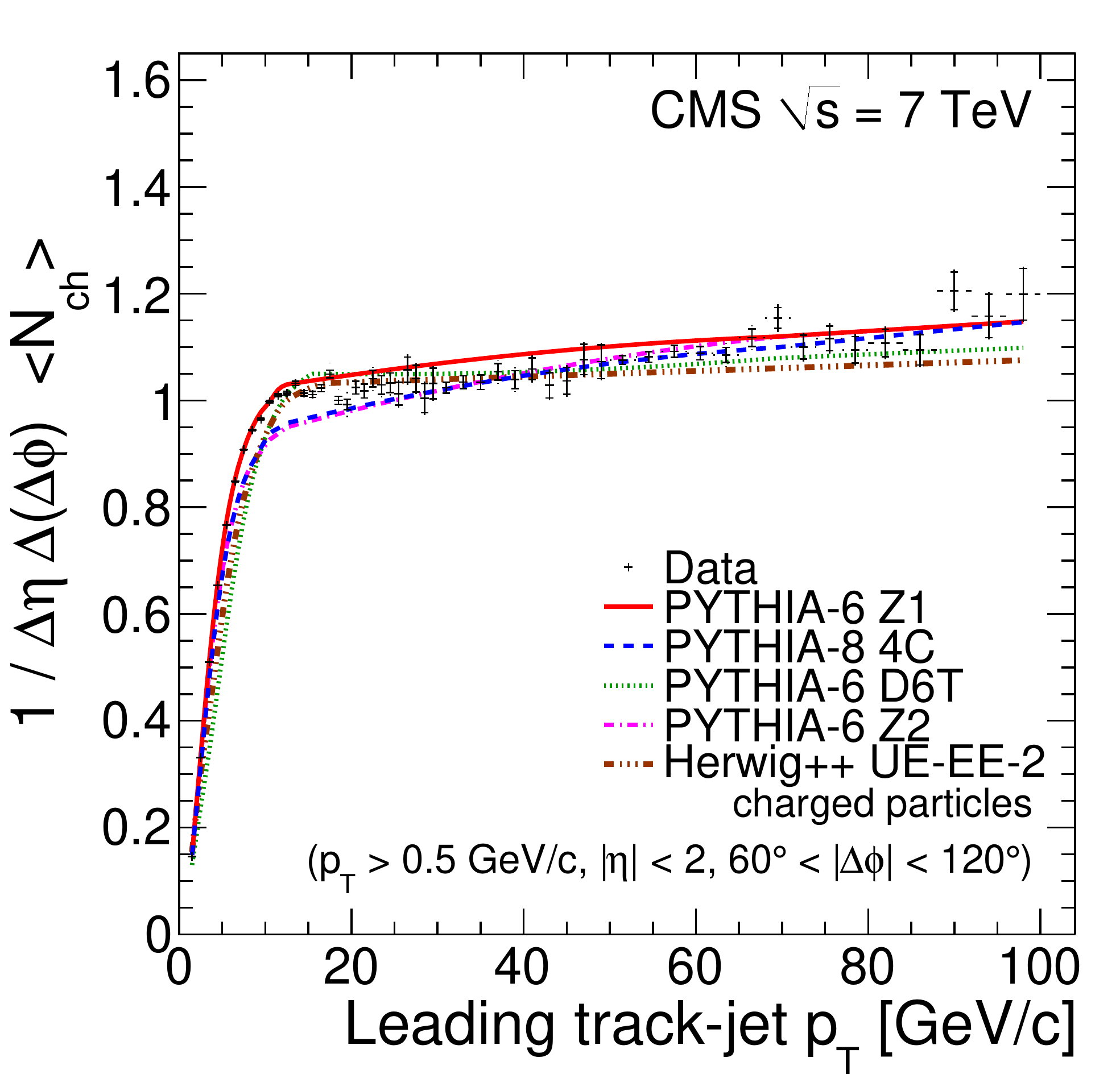}      
  \includegraphics[width=0.32\textwidth]{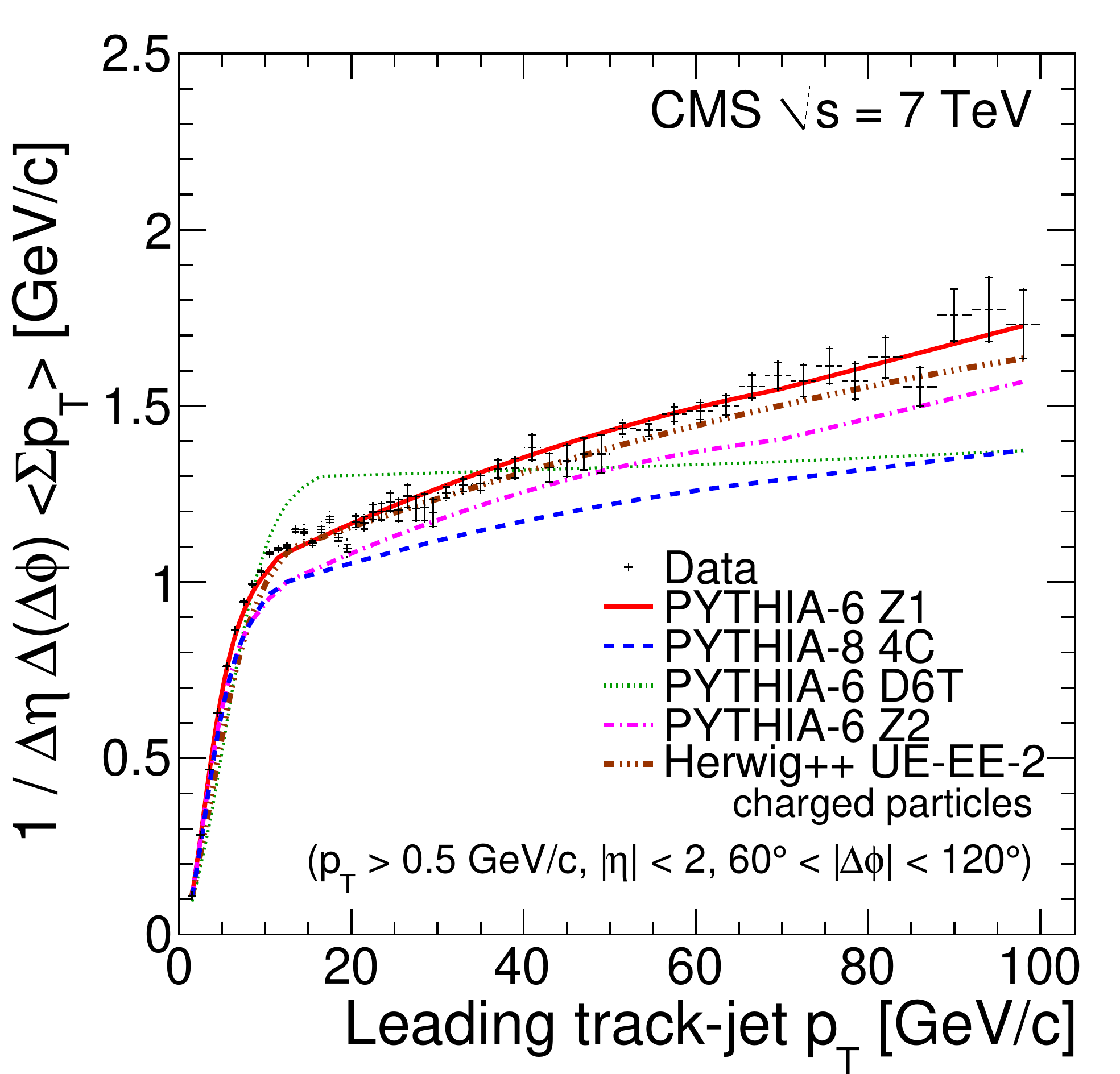}      
  \includegraphics[width=0.32\textwidth]{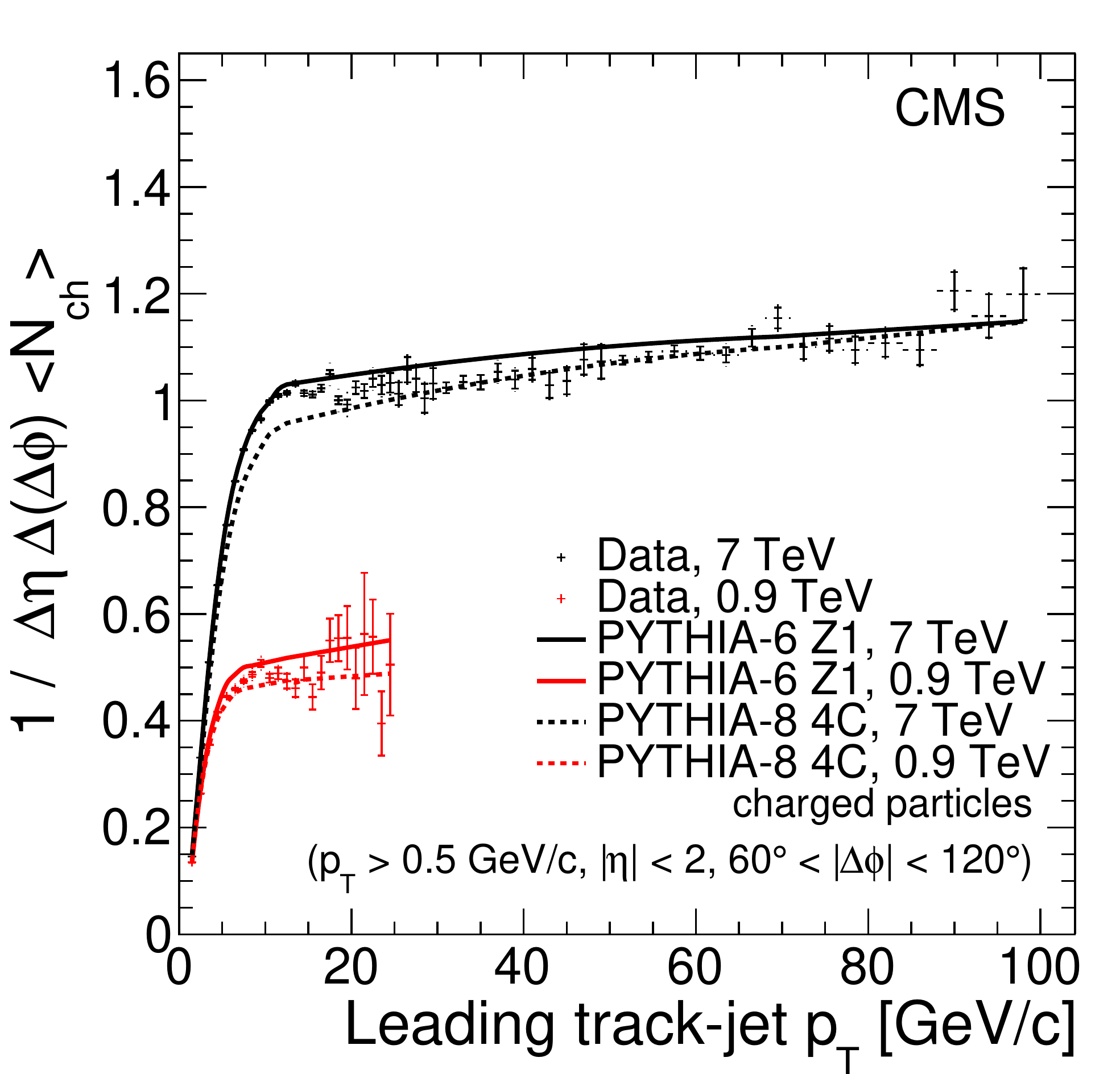}      
  \caption{Fully corrected measurements of charged particles in the transverse region.}
  \label{fig:uedensities}
\end{figure}

Fig. \ref{fig:uedensities} shows the  measurements of
charged particles with $p_T > 0.5 \gevc$ and $|\eta| < 2$ in the
transverse region. The left plot shows the average multiplicity per unit
pseudorapidity and per radian; the center plot shows the average scalar
$\Sigma p_T$ per unit pseudorapidity and per radian.

The center-of-mass energy dependence of the hadronic activity in the
transverse region is also studied. The right plot of
Fig. \ref{fig:uedensities} shows the average scalar $\Sigma p_T$, per
unit pseudorapidity and per radian, as a function of the leading
track-jet $p_T$, for data at 0.9 TeV and 7 TeV; Predictions of
three Pythia tunes are compared to the data.

\begin{figure}[!hbt]
  \centering
  \includegraphics[width=0.32\textwidth]{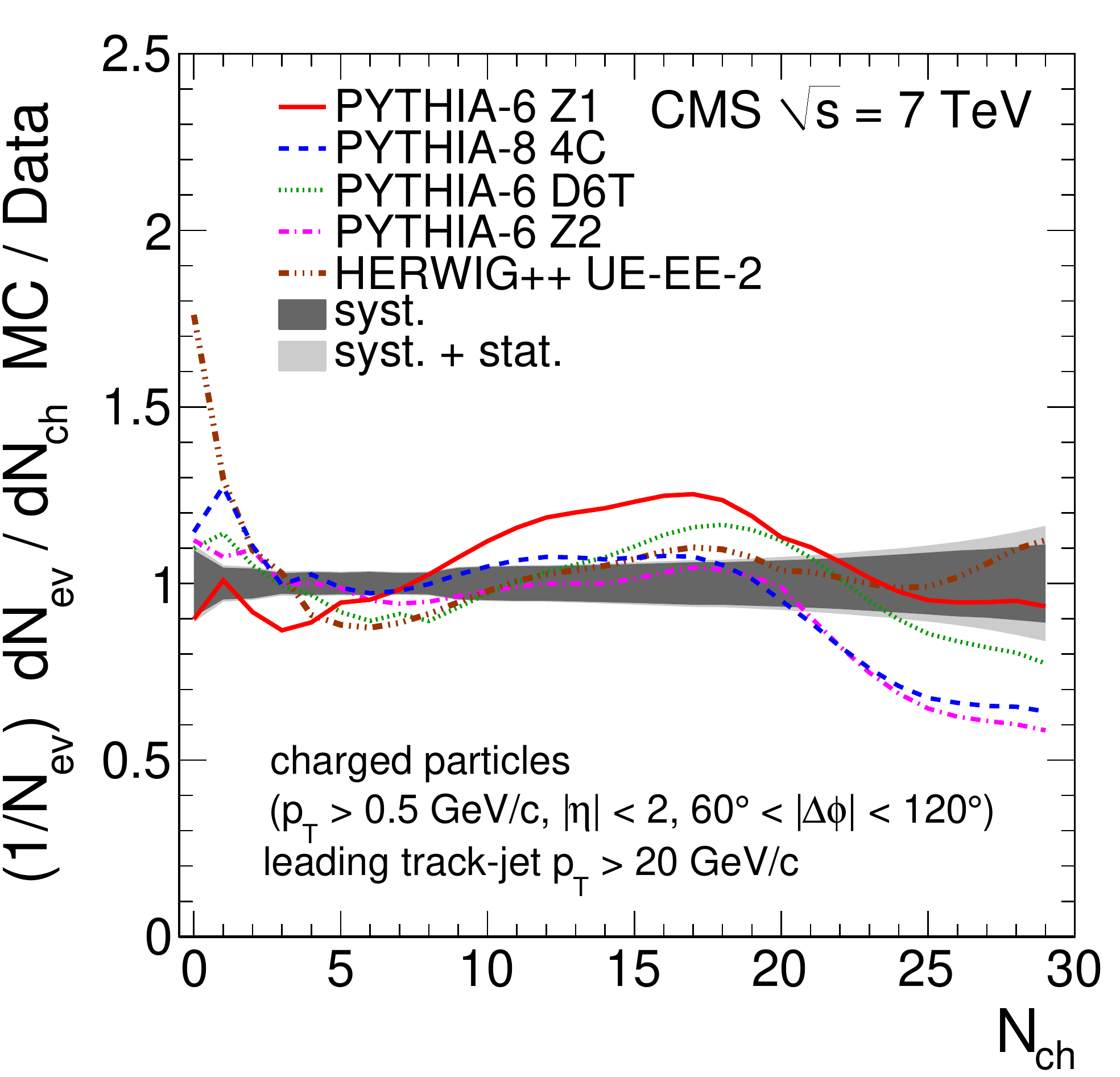}      
  \includegraphics[width=0.32\textwidth]{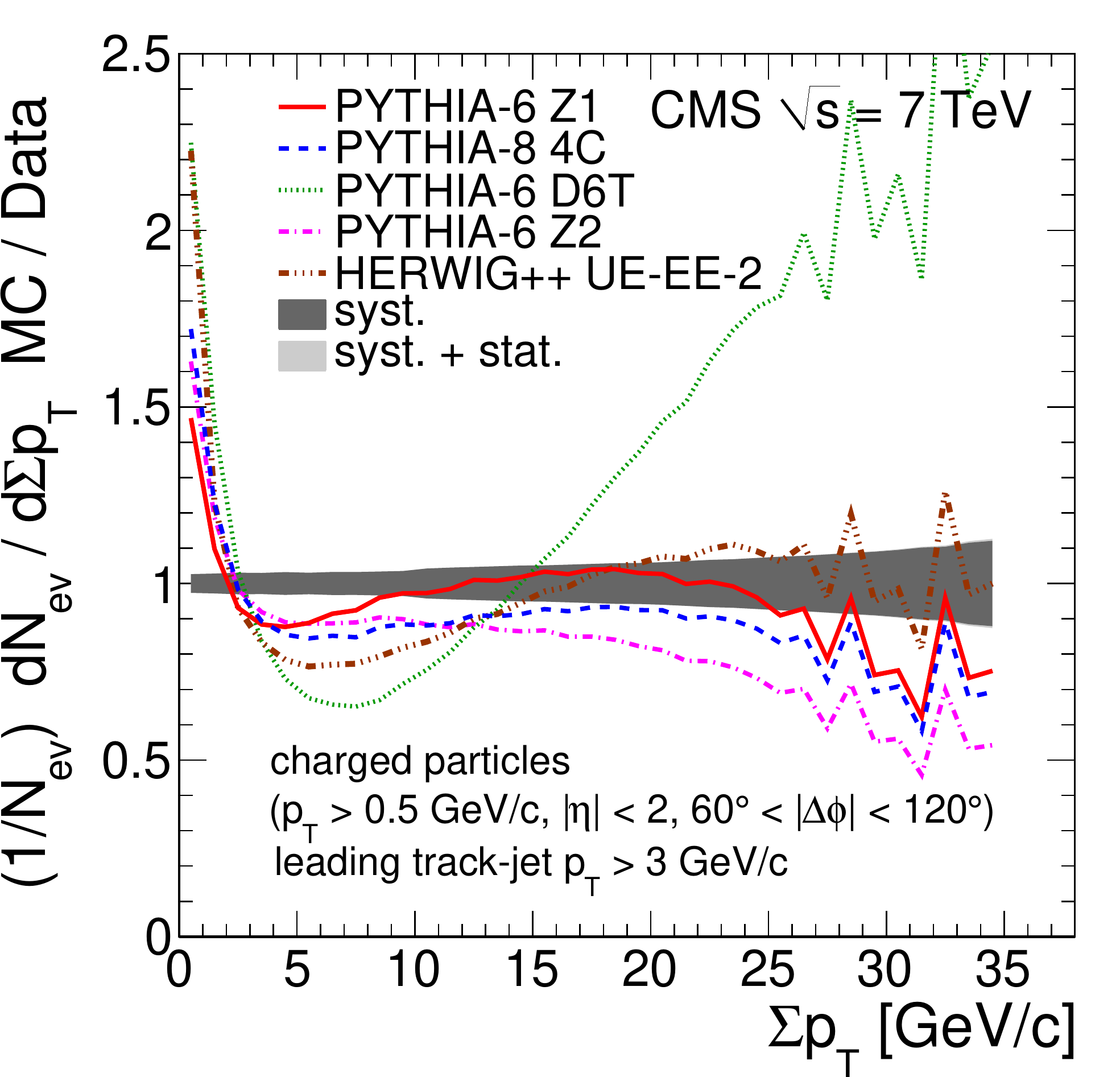}      
  \includegraphics[width=0.32\textwidth]{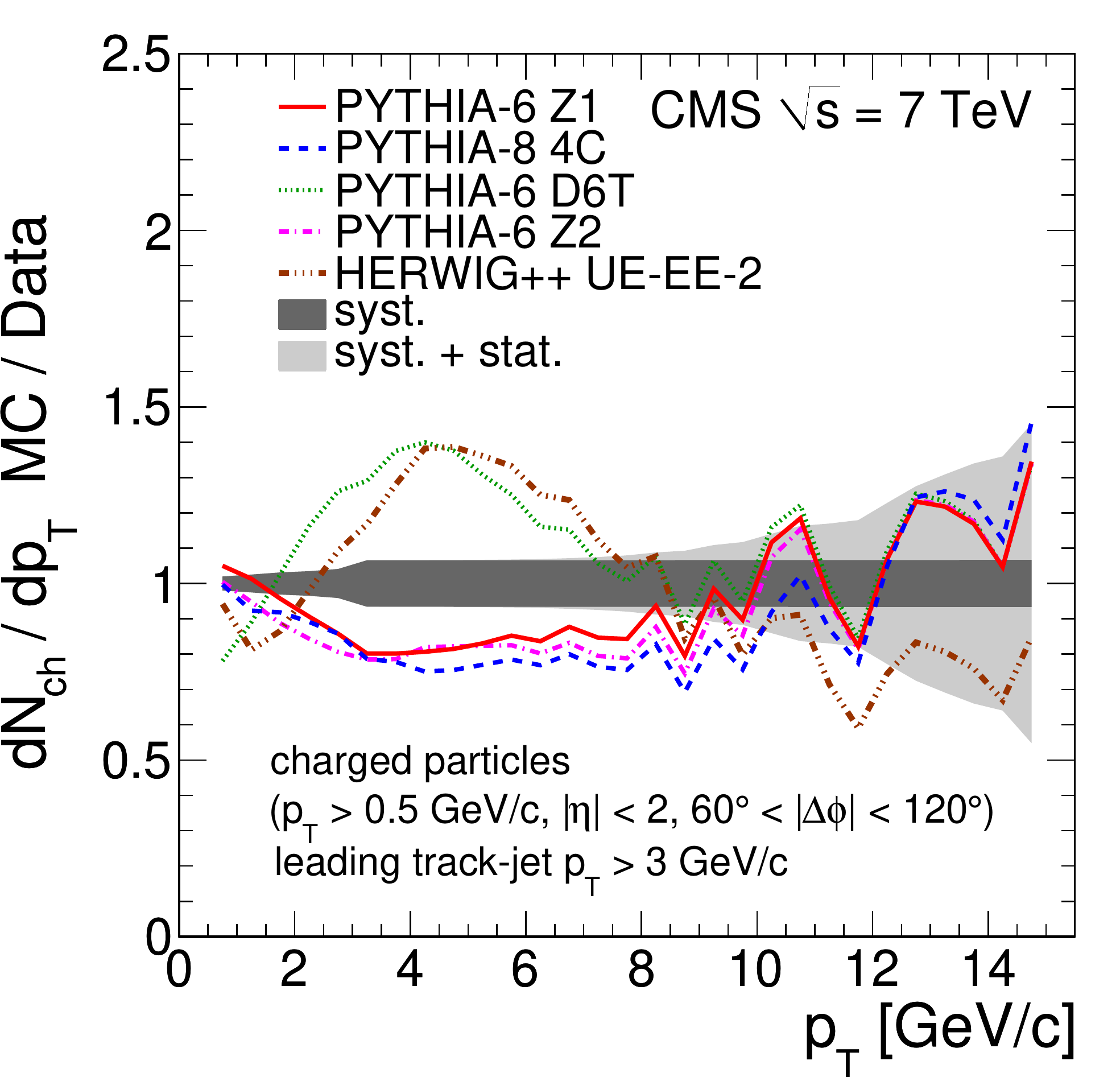}      
 
  \caption{Ratio of MC predictions to the corrected measurements of charged particles in the transverse region.}
  \label{fig:uemcdataratio}
\end{figure}

Ratios of three MC predictions to the measurements in the transverse
region are studied and shown in Fig. \ref{fig:uemcdataratio}.  The
left plot shows multiplicity distributions; the center one shows
scalar $\Sigma p_T$ distributions; the right shows the particle $p_T$
spectra.  The leading track-jet is required to have $|\eta| < 2$ and
$p_T > 3 \gevc$. The inner band corresponds to the systematic
uncertainties and the outer band corresponds to the total experimental
uncertainties. Here, statistical and systematic uncertainties are
added in quadrature.

Pythia Z1 tune turns out to describe most of these UE distributions
and the $\sqrt{s}$ dependences quite well. Tune Z2 and Pythia8
default tune 4C with new MPI model built in also do rather decent
jobs. However, from the MC/data comparisons above, there are still
cases and regions that none of them describe well enough. These CMS
measurements could thus provide important input for future tunes.

\section{Summary}
\label{sec:summary}

Measuring and understanding soft QCD processes at LHC are important
for precision Standard Model measurements and new physics
searches. CMS experiment has performed many studies in this area in pp
collisions at $\sqrt{s}$ = 0.9, 2.36 and 7 TeV at LHC.

Properties of hadron production are measured, including charged
particle transverse momentum spectra, event-by-event multiplicity
distributions and spectra of identified strange hadrons, reconstructed
based on their decay topology. Two-particle angular correlation over a
broad range of pseudorapidity and azimuthal angle in pp collisions was
measured and a pronounced structure was observed in the
two-dimensional correlation function for particle pairs with
intermediate transverse momentum of 1-3 $\gevc$ in high multiplicity
events. Bose-Einstein correlation between identical particles was
measured with respect to reference samples without
correlation. Furthermore, measurements of the underlying activities in
scattering processes with $p_T$ scale in the several GeV region were
also shown and compared to several QCD Monte Carlo models and tunes.

All these measurements have contributed new information of strong
interactions in the low transverse momenta domain, serving as inputs
to future MC modelings.



\bigskip 

\end{document}